\documentclass[12pt,journal,draftclsnofoot,letterpaper,onecolumn]{IEEEtran}

\usepackage{url}             
\usepackage[cmex10]{amsmath}
\usepackage{amsfonts}
\usepackage{amssymb}
\usepackage{cite}
\usepackage[caption=false, font=footnotesize]{subfig}

\ifCLASSINFOpdf
  \usepackage[pdftex]{graphicx}
  \DeclareGraphicsExtensions{.pdf,.jpeg,.jpg,.png}
\else
  \usepackage[dvips]{graphicx}
  \DeclareGraphicsExtensions{.eps}
\fi

\begin{document}

\title{Source and Physical-Layer Network Coding \\for Correlated Two-Way Relaying}

\author{Qiang~Huo,
        Lingyang~Song,
        Yonghui~Li,
        and~Bingli~Jiao%
\thanks{Published in IET Communications.
Manuscript received  June 15, 2015; revised October 03, 2015; accepted December 21, 2015.}%
\thanks{Q.~Huo (corresponding author) is with the Wireless Network Research Department, Huawei Technologies Co., Ltd., Shanghai 201206,
China (e-mail: qianghuoee@gmail.com).}%
\thanks{L.~Song, and B.~Jiao are with the School of Electronics Engineering and Computer Science, Peking University, Beijing 100871, China (e-mail: \{lingyang.song, jiaobl\}@pku.edu.cn).}%
\thanks{Y.~Li is with the School of Electrical and Information Engineering, the University of Sydney, Sydney, NSW 2006, Australia (e-mail: yonghui.li@sydney.edu.au).}%
\thanks{Digital Object Identifier 10.1049/iet-com.2015.0572}%
}






\maketitle


\begin{abstract}
In this paper, we study a half-duplex two-way relay channel~(TWRC) with correlated sources exchanging bidirectional information.
In the case, when both sources have the knowledge of correlation statistics, a source compression with physical-layer network coding (SCPNC) scheme is proposed
to perform the distributed compression  at each source node.
When only the relay has the knowledge of correlation statistics, we propose a relay compression with physical-layer network coding (RCPNC) scheme  to compress the bidirectional messages at the relay.
The closed-form block error rate (BLER) expressions of both schemes  are derived and verified through simulations.
It is shown that the proposed  schemes achieve
considerable improvements in both error performance and throughput compared with the conventional non-compression scheme in correlated two-way relay networks (CTWRNs).
\end{abstract}


\IEEEpeerreviewmaketitle



\section{Introduction}\label{sec:introduction}

\IEEEPARstart{N}{etwork} coding~(NC)
has been known as an efficient technique to  significantly increase the capacity of   communication networks \cite{Ahlswede2000P1204,Li2003P371,Fragouli2006P63}.
Recently, it has been applied to improve the transmission efficiency of two-way relay networks~(TWRNs)\cite{Huo2012P4998,Huo2013P5924}.
In \cite{Wu2004P,Larsson2006P851}, a three time slot~(TS) NC scheme based on decode-and-forward~(DF) protocol was first introduced to TWRNs.
By allowing two sources to transmit simultaneously to the relay,
a two TS analog network coding~(ANC) scheme with amplify-and-forward~(AF) protocol\cite{Popovski2006P588,Katti2007P397} and  a
two TS physical-layer network coding~(PNC) scheme using  estimated-and-forward (EF) protocol\cite{Zhang2006P358,Popovski2006P3885} were proposed for TWRNs.
It has been shown that exploiting NC in two-way relaying leads to substantial throughput increase with respect to the conventional four TS scheme.

However, in existing TWRNs, the information messages exchanged between two sources are assumed to be independent. In many scenarios, such as wireless sensor networks~(WSNs) \cite{Li2005Pa,Hoang2005P4,Jindal2006P466,Fang2009P3292}, surveillance systems\cite{Marcenaro2001P1419,Liu2009P453} and multiview video systems\cite{Lu2007P737,Flierl2007P66,Ding2008P1553}, the measurements exhibit spatial correlations due to the space closeness of the measurements.
Distributed source coding~(DSC) has been shown as an efficient approach to exploit this inherent correlated property\cite{Pradhan1999P158,Pradhan2000P363}.

The Slepian-Wolf~(SW) theorem shows that for the DSC problem with two statistically dependent  i.i.d.  sources $X$ and $Y$, the achievable rate
region can be bounded as $R_x\geqslant H(X|Y)$, $R_y\geqslant H(Y|X)$ and $R_x+R_y\geqslant H(X,Y)$, respectively\cite{Slepian1973P471,Cover1975P226}.
The basic idea of SW coding is to partition the space of
source sequences into bins such that
the sequences in each bin can be uniquely distinguished at the decoder with the help of side information\cite[Sec. 15.4]{Cover2006P}.
After binning, each source
transmits the bin index  of the codeword instead of the codeword itself. Since the length of the bin index is shorter than that of the codeword, the compression is achieved.
At the decoder, each user's  codewords can be decoded correctly by using the other user's correlated messages as side information.

Most existing DSC schemes are designed for the multiple access channels~(MACs), where multiple correlated sources transmit their messages to a single destination. In many applications, such as WSNs, the measurements are forwarded from each sensor node to the sink nodes based on multihop mesh network structures where sensors act as both the source nodes and relays. They need to exchange their measurements via other sensors and help each other to forward them to the sink nodes.
Moreover, as shown in a recent work \cite{Barros2006P155}, if sensor nodes can exchange correlated information over the wireless medium before transmitting the correlated data to the sink nodes, this can significantly increase the overall network transmission efficiency.
In these scenarios, we can model the part of the network as a TWRN with correlated sources exchanging their information\cite{Huo2015P30}.

In \cite{Lechner2011P466,Timo2011P253},   the authors  studied correlated two-way relay networks~(CTWRNs) from an information theoretic point of view, and necessary and sufficient conditions for reliable communication are given.
However, the study in \cite{Lechner2011P466,Timo2011P253} is conducted over  orthogonal  uplink channels, which prevents achieving higher spectral efficiency.
In \cite{Huo2015P30},  a compressed relaying scheme via Huffman and physical-layer network coding~(HPNC) was proposed for CTWRNs, where compression is performed only at the relay.
On the contrary, in this paper, we study CTWRNs from a practical coding theory point of view.
Furthermore, our study mainly focuses on the non-orthogonal   uplink transmission to achieve network throughput gain over the conventional orthogonal   uplink transmission.
Specifically,  we propose  two source and physical-layer network coding~(SPNC) schemes to achieve source compression for CTWRNs.
We first study the case when both source nodes have the knowledge of their mutual correlation statistics and propose a
source compression  with physical-layer network coding~(SCPNC) scheme to perform the distributed compression at each source node.
In the SCPNC scheme, both source nodes first
perform SW coding on their own messages   using the syndrome approach and transmit the compressed messages to the relay, simultaneously.
The relay  performs  PNC on the received symbols   and broadcasts them to both source nodes.
After receiving the  PNC-coded  symbols from the relay, each source decodes the other source's messages by using its own messages as side information.
For the application scenarios where only the relay has the correlation statistics,   a relay compression with physical-layer network coding~(RCPNC) scheme is proposed to compress the bidirectional messages at the relay.
In the RCPNC scheme, both source node transmit the correlated raw messages to the relay without compression, simultaneously.
The relay  performs PNC on the  received symbols, compresses the PNC-coded symbols using the syndrome approach and broadcasts the compressed messages  to both source nodes.
Closed-form block error rate~(BLER) expressions of
the proposed schemes are derived and compared with the conventional non-compression scheme\cite{Zhang2006P358}.
The analytical results are verified through simulations.
Simulation results  show that considerable improvements in both error performance and throughput can be achieved by exploiting the correlated property, compared with the conventional non-compression scheme.

The rest of the paper is organized as follows.
In Section \ref{sec:LBC-NC}, the system model is described, and two source compression schemes via SPNC are proposed for CTWRNs.
In Section \ref{sec:analysis}, the BLER performance  of the proposed schemes is analyzed.
Simulation results are presented in Section \ref{sec:simulations}.  Section \ref{sec:conclusion} concludes the paper.

\emph{\textbf{Notation}}: Matrices and vectors are denoted by bold capital letters and bold  lower-case letters, respectively.
$(\cdot)^\mathrm{T}$  and $(\cdot)^\mathrm{H}$ represents transpose  and Hermitian operations, respectively.
$\mathbf{I}_m$ is the $m \times m$ identity
matrix.
For a random vector variable $\mathbf{n}$,
$\mathbf{n}\sim\mathcal{CN}(0, \mathbf{\Omega})$ denotes a circular symmetric complex Gaussian variable with a zero mean and covariance matrix $\mathbf{\Omega}$.
$\mathbb{E}\{\cdot\}$  represents the expectation and $\mathbb{\sigma}\{\cdot\}$  denotes the standard deviation.
$\bigoplus$ represents XOR operation. $Q(x)$ is the $Q$-function, given by, $Q(x)=\frac{1}{\sqrt{2\pi}}\int_x^{\infty}e^{-t^2/2}\,\mathrm{d}t$.



%
%
%



\section{SPNC for Correlated Two-Way Relaying}\label{sec:LBC-NC}

\subsection{System Model}\label{sec:xx}

\begin{figure}[]
\centering
\graphicspath{{fig/}} 
\includegraphics[width=0.6\textwidth]{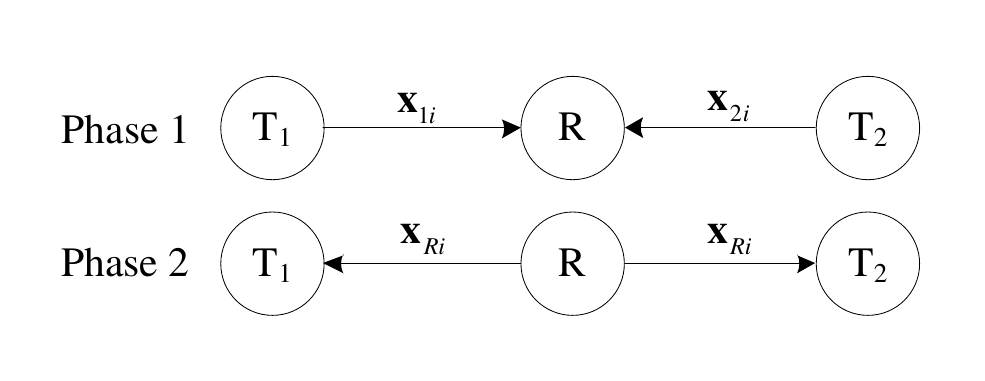}
\caption{System model for a two-hop CTWRN.}
\label{fig:TWRN_PNC}
\end{figure}

In this paper, we consider a   two-hop  CTWRN with three nodes, as shown in Fig. \ref{fig:TWRN_PNC}, where two correlated source nodes, $T_1$ and $T_2$, want to exchange correlated messages with each other through a relay node, $R$.
It is assumed that each node in the network is equipped
with one single antenna working in a half-duplex mode.
We consider an additive white Gaussian noise (AWGN) channel and binary phase-shift keying (BPSK) modulation throughout this paper.
The transmit powers of the three nodes are assumed to be equal, denoted by $\mathcal{E}$.

We assume that the message bits at both source nodes are divided equally into blocks and each block consists of $n$  bits. Also, one TS consists of $n$ symbol intervals throughput this paper.
Let $\mathbf{c}_{1i}=[c_{1i}(1), \cdots, c_{1i}(n)]$ and $\mathbf{c}_{2i}=[c_{2i}(1), \cdots, c_{2i}(n)]$ denote the $i$th-block message  of $T_1$ and $T_2$, respectively, where $\mathbf{c}_{1i}$, $\mathbf{c}_{2i} \in \{0,1\}^{1\times n}$.
The constrained source correlation model \cite{Tan2006P3102,Cao2008P1} is used in the rest of this paper. More specifically, this correlation model requires $\mathbf{c}_{1i}$ and $\mathbf{c}_{2i}$ to satisfy
\begin{equation}\label{eq:corr_model}
\begin{split}
d_H(\mathbf{c}_{1i},\mathbf{c}_{2i}) \leq t,
\end{split}
\end{equation}
where $d_H(\cdot)$ denotes the Hamming distance \cite{Lin2004P} and $t$ represents the maximum Hamming distance between two correlated source message bits within a block.  
In this paper, $t$ denotes the knowledge of correlation statistics.

\begin{figure*}[!t]
\centering
\graphicspath{{fig/}} 
\includegraphics[width=1.0\textwidth]{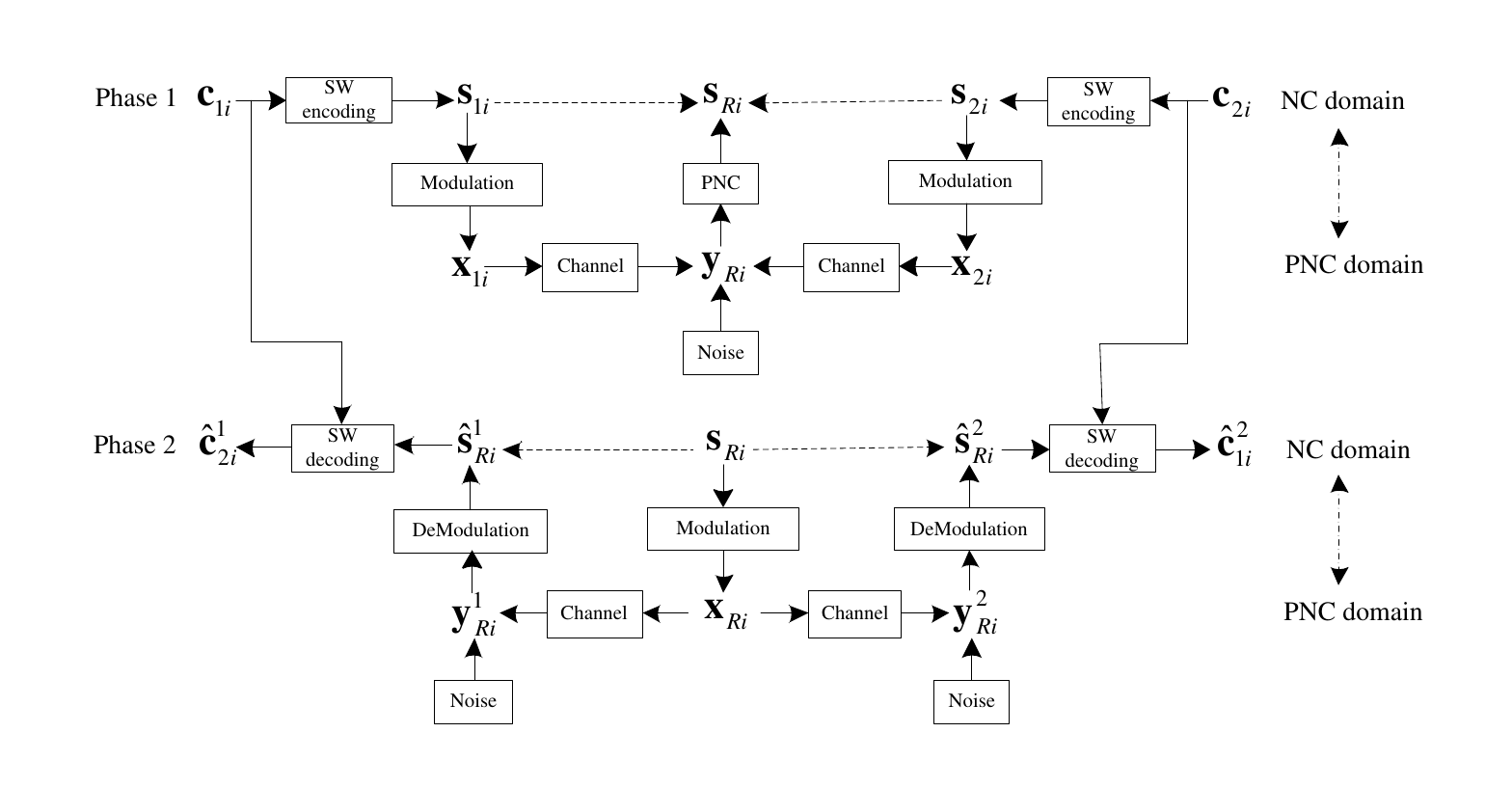}
\caption{Schematic diagram of the SCPNC scheme.}
\label{fig:TWRN_SCPNC}
\end{figure*}

\begin{figure*}[]
\centering
\graphicspath{{fig/}} 
\includegraphics[width=1.0\textwidth]{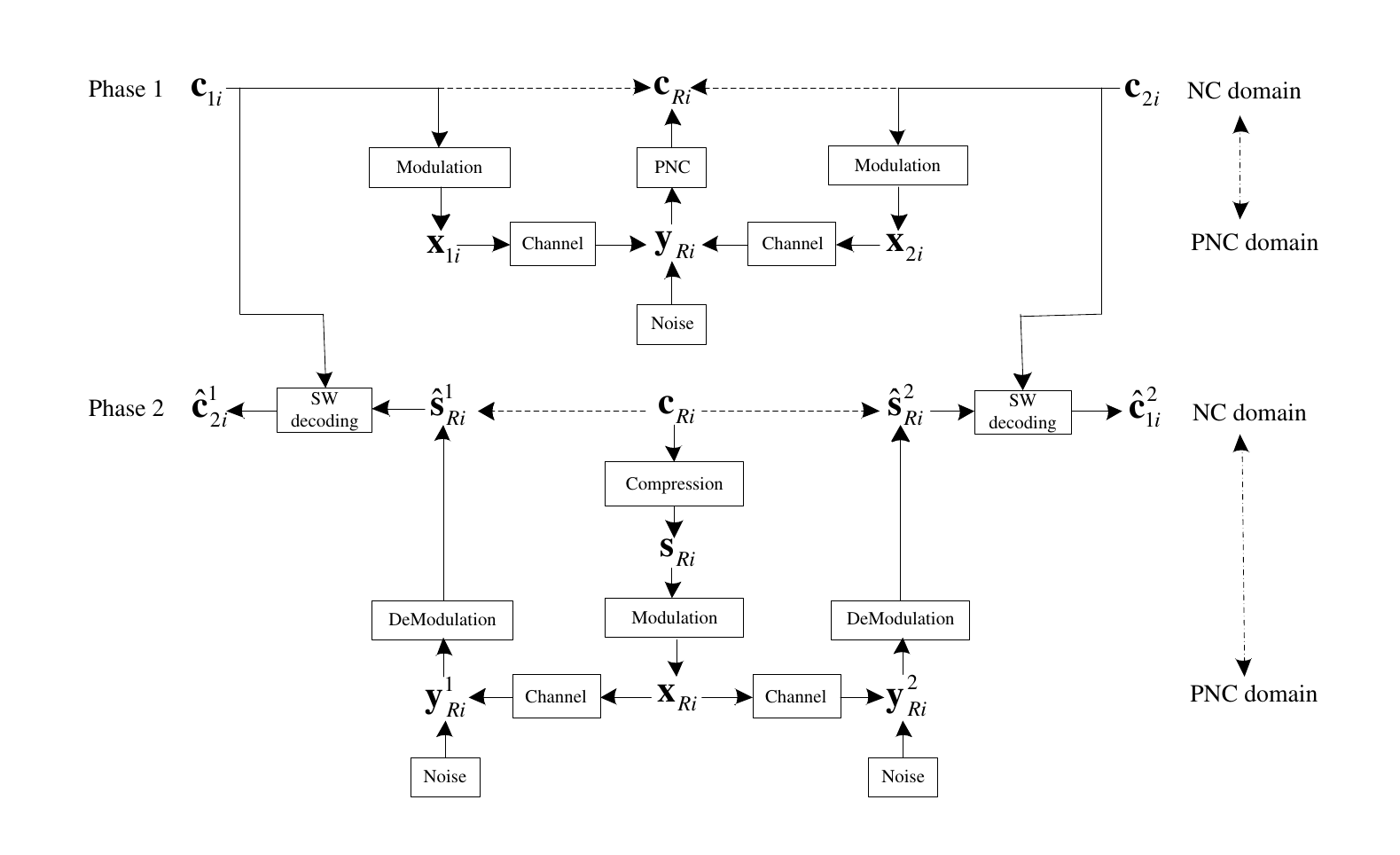}
\caption{Schematic diagram of the RCPNC scheme.}
\label{fig:TWRN_RCPNC}
\end{figure*}

In the following, we propose two source compression schemes via SPNC for CTWRNs. The corresponding schematic diagrams are shown in Fig. \ref{fig:TWRN_SCPNC} and Fig. \ref{fig:TWRN_RCPNC}, respectively.

\subsection{SCPNC Scheme: Distributed Compression at Two Source Nodes}\label{subsec:xx.yy}

\subsubsection{SW Encoding at Two Source Nodes}\label{subsec:xx.yy.zz}
The SCPNC scheme is designed for the case that both sources have  the knowledge of their correlation statistics $t$.
The compressions are performed block by block, by using a $(n,k)$ linear block code $\mathcal{C}$ with random-error-correcting capability $t$ \cite{Lin2004P}, at both source nodes independently.
We use the cosets of linear block code $\mathcal{C}$  to construct bins and each coset is indexed by its corresponding syndrome\cite{Wyner1974P2}.
In this paper, the standard array approach\cite[Sec. 3.5]{Lin2004P} is used  to construct the cosets of $\mathcal{C}$.
In such a case, each coset consists of $2^k$ vectors of length $n$.  There are totally $2^{n-k}$ isolated cosets and   every $n$-tuple  vectors  appears in one and only one coset.
Let $\mathbf{H}\in \{0,1\}^{m\times n}$ denote the parity-check matrix of code $\mathcal{C}$, where $m=n-k$.
Then the syndrome of $\mathbf{c}$ is defined as $\mathbf{s} = \mathbf{c}\cdot \mathbf{H}^\mathrm{T}$, where $\mathbf{c}  \in \{0,1\}^{1\times n}$ and $\mathbf{s}  \in \{0,1\}^{1\times m}$.
Note that all the $2^k$ $n$-tuple vectors of a coset have the same syndrome and the syndromes for different cosets are different.
Thus each syndrome uniquely represents one coset and is thus called the coset (bin) index in this paper.
In SW coding,
each message $\mathbf{c}$ is compressed into its syndrome $\mathbf{s}$. Thus, the compression ratio is $m/n=(n-k)/n$.

Let $\mathbf{s}_{1i}=[s_{1i}(1), \cdots, s_{1i}(m)]$ and $\mathbf{s}_{2i}=[s_{2i}(1), \cdots, s_{2i}(m)]$ denote the $i$th-block  compressed message  of  $\mathbf{c}_{1i}$ and $\mathbf{c}_{2i}$ at $T_1$ and $T_2$, given by,   $\mathbf{s}_{1i} = \mathbf{c}_{1i} \cdot \mathbf{H}^\mathrm{T}$ and $\mathbf{s}_{2i} = \mathbf{c}_{2i} \cdot \mathbf{H}^\mathrm{T}$, respectively. 
Then  both source nodes
modulate $\mathbf{s}_{1i}$ and $\mathbf{s}_{2i}$ into $\mathbf{x}_{1i}$ and $\mathbf{x}_{2i}$ using BPSK modulation and then transmit them to the relay simultaneously, where $\mathbf{x}_{1i} = 1-2 \mathbf{s}_{1i}$ and $\mathbf{x}_{2i} = 1-2 \mathbf{s}_{2i}$.
The corresponding $i$th-block received signal  at the relay is given by
\begin{equation}
\mathbf{y}_{Ri} = \sqrt{\mathcal{E}}\mathbf{x}_{1i} + \sqrt{\mathcal{E}}\mathbf{x}_{2i} + \mathbf{n}_{Ri},
\end{equation}
where  $\mathbf{n}_{Ri}$  is the noise vector with $\mathbf{n}_{Ri} \sim\mathcal{CN}(0, N_0 \mathbf{I}_{m})$.
The   signal-to-noise ratio  (SNR)  is denoted as
$\gamma =\frac{\mathcal{E}}{N_0}$.

\subsubsection{PNC at the Relay Node}\label{subsec:xx.yy.zz}
The relay performs PNC on the received signal. The resulting PNC-coded message is represented by $\mathbf{s}_{Ri}=\mathbf{PNC}(\mathbf{y}_{Ri})$,  where $\mathbf{s}_{Ri}$ is the estimation of $\mathbf{s}_{1i}\bigoplus\mathbf{s}_{2i}$, and $\mathbf{PNC}(\cdot)$ denotes the PNC mapping function in\cite{Zhang2006P358}.  In the PNC mapping function, when the received signal $\mathbf{y}_{Ri}$ is less than $-\gamma _{th}$ or larger than $\gamma _{th}$, we declare $\mathbf{s}_{Ri}$  to be $0$; otherwise, $\mathbf{s}_{Ri}$  is set to be $1$, where ${\gamma _{th}} = \sqrt \varepsilon   + \frac{{\sqrt {{N_0}} }}{4}\frac{1}{{\sqrt \gamma  }}\ln \left[ {1 + \sqrt {1 - {e^{ - 8\gamma }}} } \right]$ is the optimal decision threshold. $\mathbf{s}_{Ri}$ is modulated into $\mathbf{x}_{Ri}$   and then sent  to  both $T_1$ and $T_2$ during the second phase, where  $\mathbf{x}_{Ri} = 1-2 \mathbf{s}_{Ri}$.
The corresponding $i$th-block received signal vectors at  $T_1$ and $T_2$, denoted by $\mathbf{y}_{Ri}^1$ and $\mathbf{y}_{Ri}^2$,  can be written as
\begin{equation}
\mathbf{y}_{Ri}^1 = \sqrt{\mathcal{E}}\mathbf{x}_{Ri} + \mathbf{n}_{Ri}^1
 \quad \text{and} \quad
\mathbf{y}_{Ri}^2 = \sqrt{\mathcal{E}}\mathbf{x}_{Ri} + \mathbf{n}_{Ri}^2,
\end{equation}
where  $\mathbf{n}_{Ri}^1 \sim\mathcal{CN}(0, N_0 \mathbf{I}_{m})$ and  $\mathbf{n}_{Ri}^2 \sim\mathcal{CN}(0, N_0 \mathbf{I}_{m})$ are the noise vectors experienced at $T_1$ and $T_2$, respectively.

\subsubsection{SW Decoding at the Destination Nodes}\label{subsubsec:xx.yy.zz}
After receiving the PNC-coded message from the relay, $T_1$ and $T_2$ first calculate the hard estimation of $\mathbf{s}_{Ri}$, denoted by $\hat{\mathbf{s}}_{Ri}^1$ and $\hat{\mathbf{s}}_{Ri}^2$.
$T_2$ ($T_1$)  can recover the desired message bits sent from $T_1$  ($T_2$), denoted as $\hat{\mathbf{c}}_{1i}^{2}$ ($\hat{\mathbf{c}}_{2i}^{1}$), by using its own message, i.e., $\mathbf{c}_{2i}$ ($\mathbf{c}_{1i}$), as side information. The details of the decoding process are described as follows.
Since  $T_1$ and $T_2$ are mathematically symmetrical, for simplicity, in the following, we only discuss the decoding algorithm and the performance analysis at $T_2$.

To give more insights into the decoding process,  let us first consider an ideal case that  all nodes can decode the messages correctly.
In the ideal case, we have
$\hat{\mathbf{s}}_{Ri}^2
=\mathbf{s}_{1i}\bigoplus\mathbf{s}_{2i}=(\mathbf{c}_{1i}\cdot\mathbf{H}^T)\bigoplus(\mathbf{c}_{2i}\cdot\mathbf{H}^T)
=(\mathbf{c}_{1i} \bigoplus \mathbf{c}_{2i}) \cdot\mathbf{H}^T  =\mathbf{e}_{i}\cdot\mathbf{H}^T $, where  $\mathbf{e}_{i}=\mathbf{c}_{1i} \bigoplus \mathbf{c}_{2i}$ is defined as the Hamming vector.
According to the constrained source correlation model, we have $w_H(\mathbf{e}_{i})=d_H(\mathbf{c}_{1i},\mathbf{c}_{2i}) \leq t$, where $w_H(\cdot)$ denotes Hamming weight\cite{Lin2004P}.
Since $\mathbf{e}_{i}$ is within the error correcting capability of the $(n, k)$ linear block code $\mathcal{C}$, $T_2$ can
perfectly construct $\mathbf{e}_{i}$  by decoding $\hat{\mathbf{s}}_{Ri}^2$.
Finally, the desired message bits $\mathbf{c}_{1i}$ can be perfectly recovered at source $T_2$ as $\mathbf{c}_{1i}=\mathbf{e}_{i}\bigoplus \mathbf{c}_{2i}$.
  
The decoding process can be similarly extended to the non-ideal case when taking into account decoding errors at the relay and destination.
Similar to the ideal case, $T_2$  first  constructs  the  estimated Hamming vector  $\hat{\mathbf{e}}_{i}^{2}$ by decoding the  message bits received from the relay node, i.e., $\hat{\mathbf{s}}_{Ri}^2$.
Then, the estimated message bits sent from  $T_1$ are constructed as
$\hat{\mathbf{c}}_{1i}^{2}=\hat{\mathbf{e}}_{i}^{2}\bigoplus\mathbf{c}_{2i}$.

\subsection{RCPNC Scheme: Compression at the Relay}\label{subsec:xx.yy}
The RCPNC scheme is designed for the
case that only the relay has the correlation statistics $t$ and the compression is performed at the relay.
In the RCPNC scheme, both source  nodes first modulate the un-compressed raw message $\mathbf{c}_{1i}$ and $\mathbf{c}_{2i}$ into $\mathbf{x}_{1i}$ and $\mathbf{x}_{2i}$ using BPSK modulation, where $\mathbf{x}_{1i} = 1-2 \mathbf{c}_{1i}$ and $\mathbf{x}_{2i} = 1-2 \mathbf{s}_{ci}$, and then transmit them to the relay simultaneously during the first phase.
Upon receiving signal from two source nodes, denoted by $\mathbf{y}_{Ri}$, the relay performs PNC on the received signal.
The resulting PNC-coded message is represented by $\mathbf{c}_{Ri}=\mathbf{PNC}(\mathbf{y}_{Ri})$,  where $\mathbf{c}_{Ri}$ is the estimation of $\mathbf{c}_{1i}\bigoplus\mathbf{c}_{2i}$, and $\mathbf{PNC}(\cdot)$ is the PNC mapping function in\cite{Zhang2006P358}.
Then the relay compresses the PNC-coded message vector using the linear block code $\mathcal{C}$, as  $\mathbf{s}_{Ri} = \mathbf{c}_{Ri} \cdot \mathbf{H}^\mathrm{T}$.
Finally the relay modulates $\mathbf{s}_{Ri}$ into $\mathbf{x}_{Ri}$ using BPSK modulation and then sends it back to  both $T_1$ and $T_2$ during the second phase,  where $\mathbf{x}_{Ri} = 1-2 \mathbf{s}_{Ri}$.
The decoding algorithm at the destination nodes is the same as the SCPNC scheme, thus   omitted here for brevity.

\subsection{Discussion}\label{subsec:xx.yy}
Although the proposed SPNC schemes are designed for equal power allocation and AWGN channels,  they can be easily extended to unequal power allocation or general fading channels by redesigning the PNC mapping function. The interested reader may refer to \cite{Liew2013P4} for the design details on the PNC mapping function. We also refer the interested reader to \cite{Lu2013P74} for implementation details on the PNC scheme. Compared with the conventional PNC-only scheme \cite{Zhang2006P358}, extra implementation complexity for the source compression encoding and decoding is required in the proposed SPNC schemes, however, as a return, considerable improvements in both error performance and throughput are achieved.



\section{Performance Analysis}\label{sec:analysis}
In this section, the BLER of the SCPNC scheme is derived, and then extended to the RCPNC scheme  and the conventional non-compression scheme\cite{Zhang2006P358}.

We first calculate the error probabilities of  the first- and second phase transmissions of the SCPNC scheme in CTWRNs, and use these error probabilities to derive the BLER  of the SCPNC scheme.
Let  $P_\text{BPSK}(\gamma)=Q(\sqrt{2\gamma})$ denote the symbol error rate~(SER) of the BPSK system in AWGN channel\cite{Proakis2001P}, where $\gamma$ denotes the SNR and $Q(\cdot)$ is the $Q$-function.
In Appendix \ref{appendix:proof.PNC}, the  error probability  of the  PNC mapping  over single symbol is derived in a closed-form expression as $P_{\text{PNC}}(\gamma) = Q\left( \sqrt{2\gamma} + \Delta \right) + {1 \over 2} Q\left( \sqrt{2\gamma} - \Delta\right)
 - {1 \over 2}Q\left( 3\sqrt{2\gamma} + \Delta\right)  $,  where $\Delta ={{\sqrt 2 } \over 4 \sqrt \gamma} \ln \left[ {1 + \sqrt {1 - {e^{ - 8\gamma }}} }\right]$.
According to \cite{David2003P}, the error probability  of  the PNC mapping during the first phase  can be expressed as
\begin{equation}\label{eq:T1_R_BLER}
\begin{split}
  P_{R}(\gamma)   = 1-\left(1-P_{\text{PNC}}(\gamma)\right)^{n-k}.\\
\end{split}
\end{equation}
The error probability  of the transmission of  $\mathbf{s}_{Ri}$ from $R$ to $T_2$  during the second phase  is given by
\begin{equation}\label{eq:R_T2_BLER}
\begin{split}
  P_{R2}(\gamma)  = 1-(1-P_\text{BPSK}(\gamma))^{n-k}.\\
\end{split}
\end{equation}

Let $P_{12}^{\text{SCPNC}}$ denote the BLER of the transmission from $T_1$ to $T_2$  in the SCPNC scheme.
Under the assumption of the constrained source correlation model\cite{Tan2006P3102}, i.e., Eq. (\ref{eq:corr_model}), a block error, i.e., $\hat{\mathbf{c}}_{1i}^{2} \neq \mathbf{c}_{1i}$, only occurs if the transmission of the corresponding syndrome fails, i.e., $\hat{\mathbf{s}}_{Ri}^2 \neq \mathbf{s}_{1i}\bigoplus\mathbf{s}_{2i}$.
Thus, $P_{12}^{\text{SCPNC}}$ can be calculated as\cite{David2003P}
\begin{equation}\label{eq:T1_T2_BLER}
\begin{split}
   & P_{12}^{\text{SCPNC}}(\gamma)
     =  1-\left(1-P_{R}(\gamma)\right)   \left(1-P_{R2}(\gamma)\right). \\
\end{split}
\end{equation}
In Appendix \ref{appendix:proof.PNC}, we have $P_{\text{PNC}}(\gamma)\approx \frac{3}{2}Q( {\sqrt {2\gamma}})$ at high SNR regime.
Applying the approximation $1-(1-x)^{N}\approx N x$ when $x$ is small,  at high SNR regime, Eq. (\ref{eq:T1_T2_BLER})  can be approximated as
\begin{equation}\label{eq:T2_T1_BLER_approx}
\begin{split}
  P_{12}^{\text{SCPNC}}(\gamma)
     \approx & P_{R}(\gamma)  +  P_{R2}(\gamma)
     \approx  \frac{5(n-k)}{2}Q(\sqrt{2\gamma}).
\end{split}
\end{equation}

Similarly, the asymptotic  BLER expressions of the RCPNC scheme and the conventional non-compression scheme are derived as
\begin{equation}\label{eq:T2_T1_BLER_approx_S2}
\begin{split}
  P_{12}^{\text{RCPNC}}(\gamma)
     \approx & \frac{5n-2k}{2}Q(\sqrt{2\gamma}),
\end{split}
\end{equation}
and
\begin{equation}\label{eq:T2_T1_BLER_approx_trad}
\begin{split}
   P_{12}^{\text{Conv}}(\gamma)
     \approx & \frac{5n}{2} Q(\sqrt{2\gamma}),
\end{split}
\end{equation}
respectively.
The exact BLER expressions of these two cases are omitted here due to the limited space.

Comparing Eq. (\ref{eq:T2_T1_BLER_approx}) and Eq. (\ref{eq:T2_T1_BLER_approx_S2}) with Eq. (\ref{eq:T2_T1_BLER_approx_trad}),
  the gain of the proposed schemes over the conventional scheme in terms of  the ratio of BLER at high SNR regime yields
\begin{equation} \label{eq:BLER_gain_S1}
\begin{split}
G_{\text{BLER}}^{\text{SCPNC}}=\lim_{\gamma \rightarrow\infty}\frac{P_{12}^{\text{Conv}}(\gamma)}{P_{12}^{\text{SCPNC}}(\gamma)} = \frac{n}{n-k},
\end{split}
\end{equation}
and
\begin{equation} \label{eq:BLER_gain_S2}
\begin{split}
G_{\text{BLER}}^{\text{RCPNC}}=\lim_{\gamma \rightarrow\infty}\frac{P_{12}^{\text{Conv}}(\gamma)}{P_{12}^{\text{RCPNC}}(\gamma)} = \frac{5n}{5n-2k},
\end{split}
\end{equation}
respectively.  Eq. (\ref{eq:BLER_gain_S1})    shows  that the  proposed SCPNC scheme has superior BLER performance compared with the conventional scheme, while Eq. (\ref{eq:BLER_gain_S2})
shows that the RCPNC scheme also outperforms the conventional scheme, but with a smaller BLER performance gain. We should emphasize that Eq. (\ref{eq:BLER_gain_S1}) and Eq. (\ref{eq:BLER_gain_S2}) are only derived for high SNR regime and the performance gains are due to  the compression of the bi-directional messages.



\section{Simulation Reuslts}\label{sec:simulations}

\begin{figure}[]
\centering
\graphicspath{{fig/}} 
\includegraphics[width=0.75\textwidth]{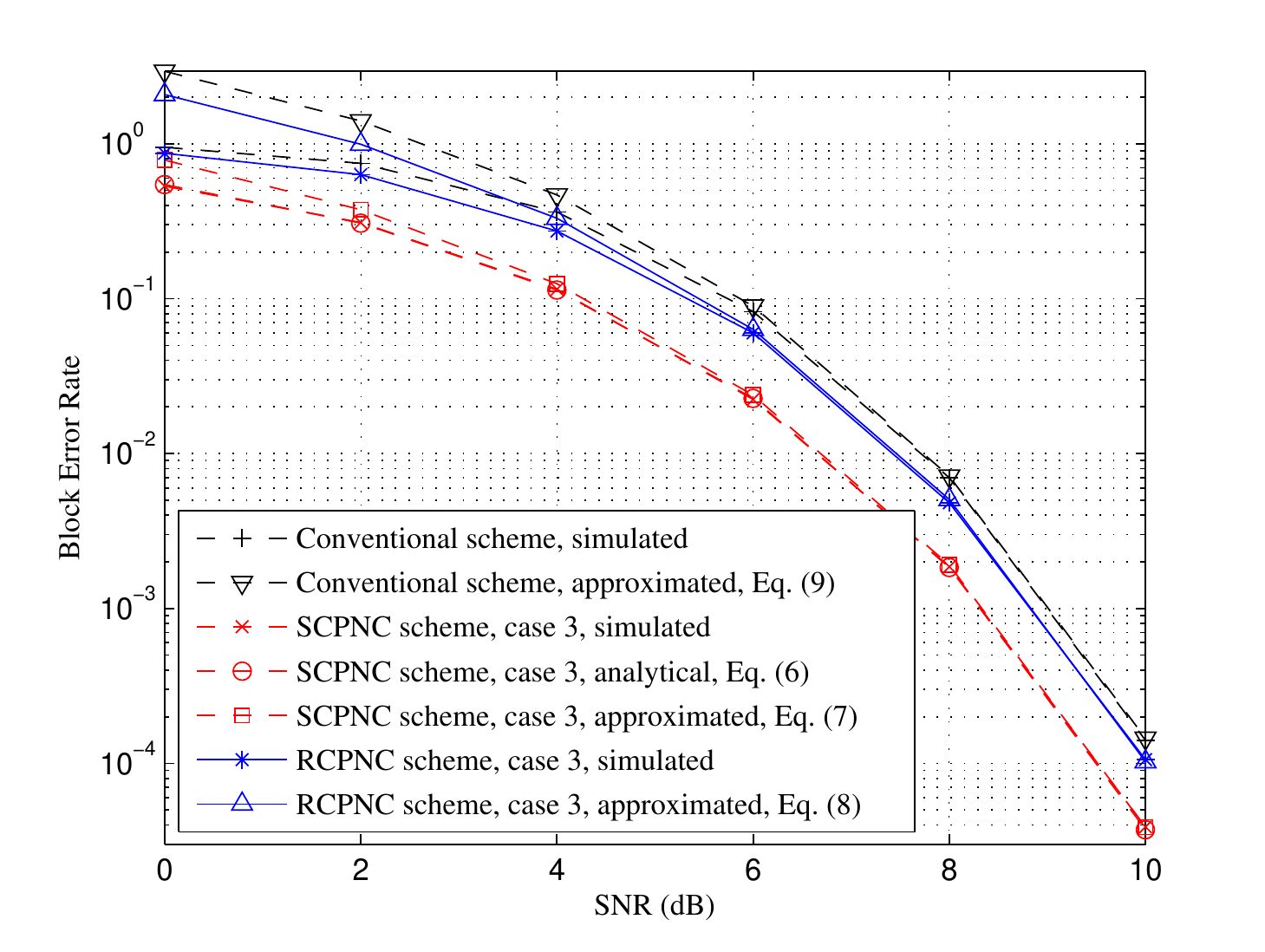}
\caption{BLER performance of the SPNC schemes and the conventional scheme.}
\label{fig:corr_LBC-PNC_BLER_case3}
\end{figure}

In this section, we provide analytical and simulated results for the proposed SPNC  schemes in CTWRNs.
Simulations  are performed with  BPSK modulation over AWGN channels. Each block consists of $n=15$ bits and   the constrained source correlation model \cite{Tan2006P3102,Cao2008P1} is used.
We consider the following three cases: (1) $t=3$, $\rho_1=60\%$; (2) $t=2$, $\rho_2 =73.33\%$; and (3) $t=1$, $\rho_3 =86.67\%$, where  $\rho_1$, $\rho_2$, and $\rho_3$ are the correlation factors which are defined as
$\rho =\frac{\min\left\{|(\mathbf{c}_{1i}-\mathbb{E}\{\mathbf{c}_{1i}\})(\mathbf{c}_{2i}-\mathbb{E}\{\mathbf{c}_{2i}\})^\mathrm{H}|\right\}}{\mathbb{\sigma}\{\mathbf{c}_{1i}\}\mathbb{\sigma}\{\mathbf{c}_{2i}\}^\mathrm{H}}$\cite{Zwillinger2000P}. We use BCH codes \cite{Bose1960P68} as our compression linear block codes and select the codes for the above three cases as:
(1) $(15,5)$ BCH code for $t=3$; (2) $(15,7)$ BCH code for $t=2$; and (3) $(15,11)$ BCH code for $t=1$.
The compression rates of the three cases are $2/3$, $8/15$ and $4/15$, respectively.

\begin{figure}[]
\centering
\graphicspath{{fig/}} 
\includegraphics[width=0.75\textwidth]{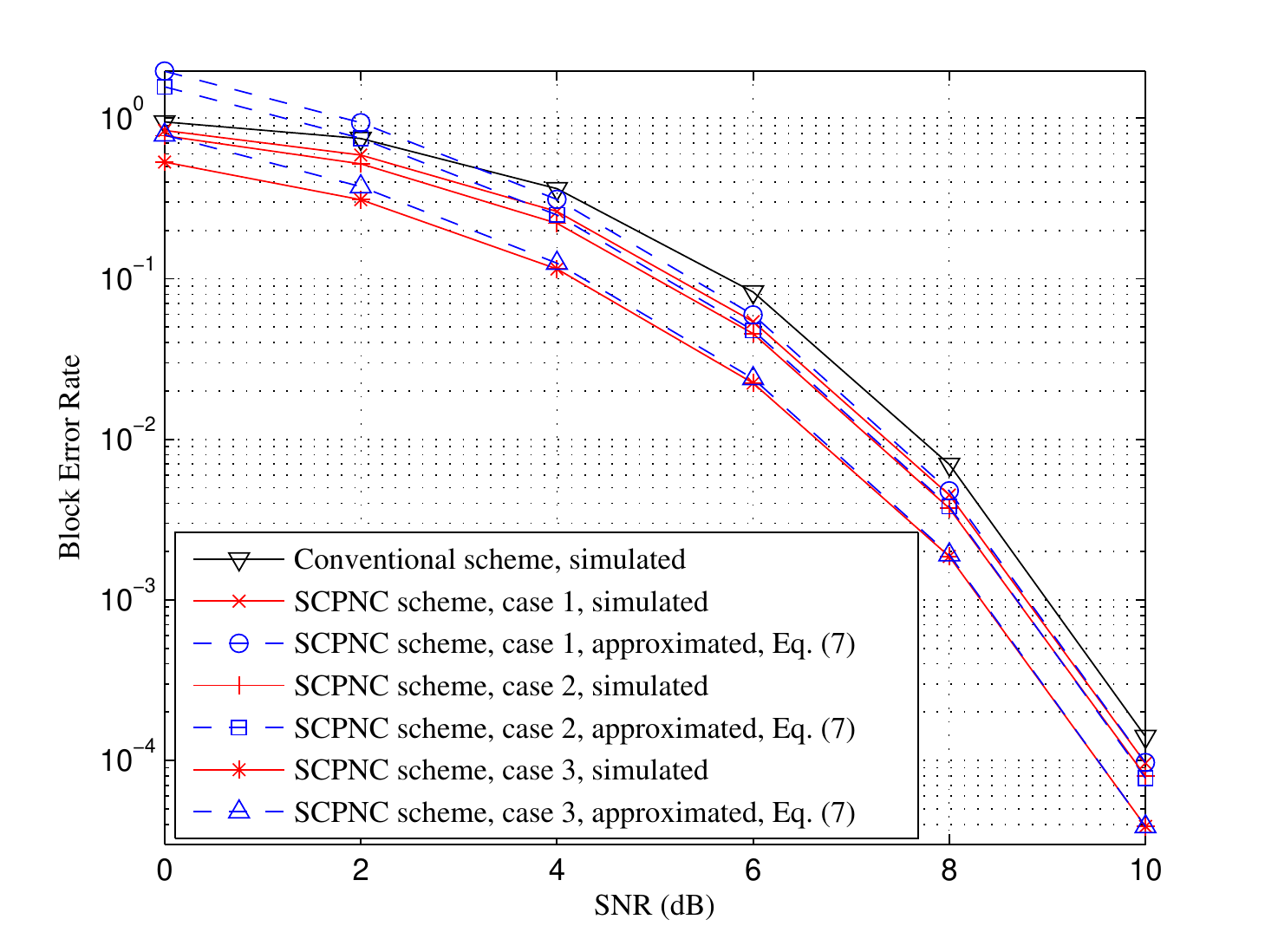}
\caption{BLER performance of the SCPNC scheme and the conventional scheme.}
\label{fig:corr_LBC-PNC_BLER_case123}
\end{figure}

\begin{figure}[]
\centering
\graphicspath{{fig/}}
\includegraphics[width=0.75\textwidth]{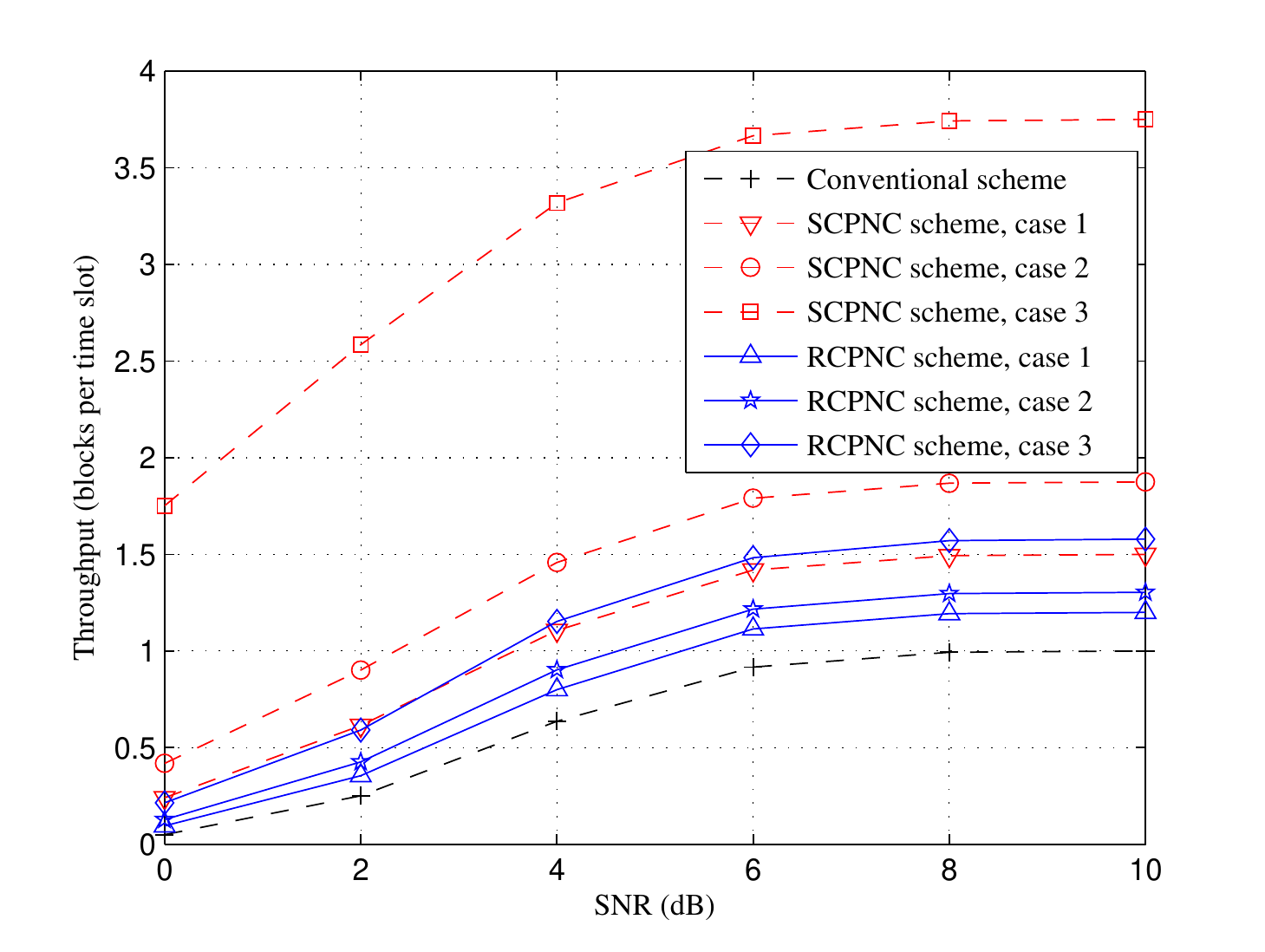}
\caption{Throughput comparison between the SPNC schemes and the conventional scheme.}
\label{fig:corr_LBC-NC_throughput}
\end{figure}

Fig. \ref{fig:corr_LBC-PNC_BLER_case3} shows the  analytical and simulated BLER performance of the SPNC schemes and the conventional scheme.
It can be observed that the closed-form expression in Eq. (\ref{eq:T1_T2_BLER}) matches perfectly with the simulated result, and the asymptotic expressions in Eq. (\ref{eq:T2_T1_BLER_approx}), Eq. (\ref{eq:T2_T1_BLER_approx_S2}) and Eq. (\ref{eq:T2_T1_BLER_approx_trad}) are also very accurate at medium to high SNR regime.   This validates the accuracy of the  BLER analysis.
In Fig. \ref{fig:corr_LBC-PNC_BLER_case3}, we also compare the BLER performance of the proposed SPNC schemes and the conventional scheme.
It can be observed that the  proposed SCPNC scheme considerably outperforms the conventional non-compression scheme.
Fig. \ref{fig:corr_LBC-PNC_BLER_case3} also shows
that the RCPNC scheme also outperforms the conventional scheme, but the gain brought by the RCPNC scheme is smaller compared with the SCPNC scheme.
This is consistent with the analytical result in Section \ref{sec:analysis}.
Also, Fig. \ref{fig:corr_LBC-PNC_BLER_case123} shows that the performance gain increases substantially  as the correlation factor increases in the  proposed SCPNC scheme.

In Fig. \ref{fig:corr_LBC-NC_throughput}, we present the simulated throughput for the SPNC schemes and the conventional scheme.
The throughput is defined as the number of  message blocks ($\mathbf{c}_{2i}$ and $\mathbf{c}_{1i}$) which are decoded correctly at source node $T_1$ and $T_2$ per TS, where one TS consists of $n$ symbol intervals.
It shows that the   proposed SCPNC scheme  achieves significant throughput improvement
with respect to the conventional scheme at the whole SNR regime and the RCPNC scheme can also bring considerable throughput gain, which is smaller than the SCPNC scheme.  It is shown that the throughput increases substantially  as the correlation factor increases in both  SPNC schemes.



\section{Conclusion }\label{sec:conclusion}

In this paper, two source compression schemes via SPNC were proposed for CTWRNs.
Analytical BLER expressions are derived   for the proposed  schemes and verified by simulations.
It has been shown that the proposed schemes  achieve  substantial improvements in both BLER performance and throughput  compared with the conventional non-compression scheme.
The improvements are contributed from the compression of the bidirectional messages.



\appendices


\section{}\label{appendix:proof.PNC}
According to \cite{Zhang2006P358}, the  error probability  of the  PNC mapping  over single symbol can be expressed in a integral form  as
\begin{equation} \label{eq:PNC-error}
\begin{split}
P_{\text{PNC}}(\gamma)&= \frac{1}{2}\int_{ - \infty }^{ - {\gamma _{th}}} {\frac{1}{{\sqrt {\pi {N_0}} }}} \exp \left( {\frac{{ - {t^2}}}{{{N_0}}}} \right)dt \\
&\qquad + \frac{1}{2}\int_{{\gamma _{th}}}^\infty  {\frac{1}{{\sqrt {\pi {N_0}} }}} \exp \left( {\frac{{ - {t^2}}}{{{N_0}}}} \right)dt \\
&\qquad  + \frac{1}{4}\int_{ - {\gamma _{th}}}^{{\gamma _{th}}} {\frac{1}{{\sqrt {\pi {N_0}} }}} \exp \left( {\frac{{ - {{\left( {t + 2\sqrt \varepsilon  } \right)}^2}}}{{{N_0}}}} \right)dt \\
&\qquad + \frac{1}{4}\int_{ - {\gamma _{th}}}^{{\gamma _{th}}} {\frac{1}{{\sqrt {\pi {N_0}} }}} \exp \left( {\frac{{ - {{\left( {t - 2\sqrt \varepsilon  } \right)}^2}}}{{{N_0}}}} \right)dt,
\end{split}
\end{equation}
where ${\gamma _{th}} = \sqrt \varepsilon   + \frac{{\sqrt {{N_0}} }}{4}\frac{1}{{\sqrt \gamma  }}\ln \left[ {1 + \sqrt {1 - {e^{ - 8\gamma }}} } \right]$ is the optimal decision threshold. After some mathematical manipulations and by using the  $Q$-function, Eq. (\ref{eq:PNC-error}) can be rewritten in a closed-form expression as\cite{Huo2015P30}
\begin{equation} \label{eq:PNC-error-2}
\begin{split}
P_{\text{PNC}}(\gamma) &= Q\left( {{{\bar \gamma }_{th}}} \right) + {1 \over 2} Q\left( { - {{\bar \gamma }_{th}} + 2\sqrt {2\gamma } } \right) \\
&\qquad - {1 \over 2}Q\left( {{{\bar \gamma }_{th}} + 2\sqrt {2 \gamma} } \right)\\
&= Q\left( \sqrt{2\gamma} + \Delta \right) + {1 \over 2} Q\left( \sqrt{2\gamma} - \Delta\right) \\
&\qquad - {1 \over 2}Q\left( 3\sqrt{2\gamma} + \Delta\right),
\end{split}
\end{equation}
where ${\bar \gamma _{th}} =  \sqrt{2\gamma}  + {{\sqrt 2 } \over 4 \sqrt \gamma} \ln \left[ {1 + \sqrt {1 - {e^{ - 8\gamma }}} } \right]=  \sqrt{2\gamma}  + \Delta$ and $\Delta ={{\sqrt 2 } \over 4 \sqrt \gamma} \ln \left[ {1 + \sqrt {1 - {e^{ - 8\gamma }}} }\right]$.
At high SNR, we have $\lim_{\gamma  \to \infty } \Delta = 0$, $Q\left( \sqrt{2\gamma} + \Delta \right) \approx Q\left( \sqrt{2\gamma}  \right)$ and
$Q\left( \sqrt{2\gamma} - \Delta\right)-\left( 3\sqrt{2\gamma} + \Delta\right) \approx  Q\left( \sqrt{2\gamma}\right)$. Thus, at high SNR regime, Eq. (\ref{eq:PNC-error-2}) can be approximated as
\begin{equation} \label{eq:PNC-error-3}
\begin{split}
P_{\text{PNC}}(\gamma)\approx \frac{3}{2}Q( {\sqrt {2\gamma}}).
\end{split}
\end{equation}





%




\bibliographystyle{IEEEtran}
\bibliography{IEEEabrv,Corr.SW.LBC_PNC}

%
%

















\end{document}